# Stability and Stabilization of Fractional-order Systems with Different Derivative Orders: An LMI Approach

Pouya Badri[1], Mahdi Sojoodi[1*]

[1]Advanced Control Systems Laboratory, School of Electrical and Computer Engineering, Tarbiat Modares University, Tehran, Iran.

*Abstract—* **Stability and stabilization analysis of fractional-order linear time-invariant (FO-LTI) systems with different derivative orders is studied in this paper. First, by using an appropriate linear matrix function, a single-order equivalent system for the given different-order system is introduced by which a new stability condition is obtained that is easier to check in practice than the conditions known up to now. Then the stabilization problem of fractional-order linear systems with different fractional orders via a dynamic output feedback controller with a predetermined order is investigated, utilizing the proposed stability criterion. The proposed stability and stabilization theorems are applicable to FO-LTI systems with different fractional orders in one or both of $0 < \alpha < 1$ and $1 \leq \alpha < 2$ intervals. Eventually, some numerical examples are presented to confirm the obtained analytical results.**

I. INTRODUCTION

Recently, study of applications of fractional calculus in the modeling and control of various real-world systems have attracted increasing interest [1–6]. This is mainly due to the fact that fractional-order models can more precisely describe systems having responses with anomalous dynamics and long memory transients. Accordingly, controller-designing methods for systems modeled by fractional-order dynamics are of great interest. Utilizing fractional-order controllers is a proper method for controlling such systems, because of their high performance and robustness [7-9]. In this paper, we are particularly concerned with fractional-order linear time-invariant (LTI) systems with different orders. This kind of system representation is used to analyze linear electrical circuits composed of resistors, supercondensators (ultracapacitors), coils, and voltage (current) sources [10], and formulate the problem of model reference adaptive control of fractional-order systems presented in [11]. Moreover, relaxation processes, viscoelastic materials models, and diffusion phenomena can be easily modeled by multi-order fractional differential equations [12]. In [2, 13, 14] non-integer order of the proposed controllers are different from the plant orders, which results in multi-order closed-loop systems.

The stability of fractional-order feedback systems was investigated in [15], by adapting classical root locus plot analysis to some viscoelastic structures. In [16], the asymptotic stability of fractional-order linear systems with delayed entry and delayed state was ensured by designing a sliding mode control. Stability of FO-LTI systems with commensurate different fractional orders is addressed in [17], using its characteristic equation. In addition, multi-term fractional differential equations such as Bagley–Torvik and Basset equations were discussed in [18], where the solution is proposed under the assumption that

The authors are with the School of Electrical and Computer Engineering, Tarbiat Modares University, Tehran, Iran, P.O. Box 14115.
Pouya.badri@modares.ac.ir, Sojoodi@modares.ac.ir (corresponding author to provide phone: +98 21 8288-3902)

fractional derivatives are rational numbers. Furthermore, stabilization of uncertain multi-order fractional systems with fractional orders $0 < \alpha < 1$ is investigated in [19] using extended state observer.

The utilization of linear matrix inequality (LMI)-based methods has been increased by the evolution of efficient numerical procedures to solve convex optimization problems [20–26] defined by LMI conditions. Appropriate LMI conditions for asymptotic stability analysis of FO-LTI systems with fractional orders in $1 < \alpha < 2$ and $0 < \alpha < 1$ intervals are presented in [27] and [28], respectively. The problem of pseudo-state feedback stabilization of fractional-order systems is presented in [28]. Necessary and sufficient conditions for the stability and stabilization of fractional-order interval systems is addressed in [29].

State feedback controller is utilized in the most of mentioned studies, in which all individual states are needed. However, in some cases measuring all states is impossible due to economic issues or physical limitations, where using output feedback control seems to be effective [20]. Moreover, it should be noted that dynamic controllers are superior to the static ones because of their more performance efficiency and degrees of freedom in achieving control objectives [30]. A large number of controller design procedures lead to high-order controllers with expensive implementation, undesirable reliability, maintenance difficulties, and potential numerical errors. Therefore, designing a controller with a low and fixed order, chosen by the designer according to the requirements of the problem, is an efficient solution to this problem [20].

Motivated by the above discussion, the main contributions of the paper are two-fold. First, using an appropriate linear matrix function, we introduce an equivalent system for the given commensurate-different-order system, by which a new stability condition is derived that is easier to check in practice comparing to the condition known so far. The presented equivalent system is single-order, in the sense that the order of non-integer derivatives of the system is the same for all pseudo-states. Second, the stabilization problem of fractional-order linear systems with different fractional orders through a dynamic output feedback controller with a predetermined order is investigated based on the proposed stability theorem. As far as we know, there is no result on the analytical design of a stabilizing dynamic output feedback controller for fractional-order systems with different orders in the literature.

The rest of this paper is organized as follows. In section II, some preliminary concepts, lemmas, and definitions are included. Problem formulation is presented in section III. Section IV is devoted to the main results of the paper where a single-order equivalent system for the given different-order system is presented, a new stability condition is investigated for such systems, and the stabilization analysis of fractional-order systems with different orders is addressed. Some numerical examples are provided in Section V to illustrate the efficiency and validity of the proposed method. Eventually, section VI draws the conclusion.

**Notation:** Throughout the paper $\uparrow$ is the symbol of pseudo inverse and $\gcd(d_1, d_2, \ldots, d_n)$ stands for the greatest common divisor of n-tuple $(d_1, d_2, \ldots, d_n)$. $A_{i,*}$ and $A_{*,j}$ represent the $i$-th row and $j$-th column of matrix $A$, respectively. Moreover, **0** is a matrix, of suitable dimensions, with all its entries set to zero and $i$ denotes the imaginary unit. By $M^T$, $\bar{M}$ and $M^*$, we denote the transpose, the conjugate, and the transpose conjugate of $M$, resepectively. Function space $C^k[a,b]$ is the space of functions $f$, having continuous $k$-th derivative. Furthermore, symbols $\mathbb{N}$, $\mathcal{R}$, $\mathcal{C}$, respectively stand for natural number, real number, and complex number set. By $\lceil \cdot \rceil$, $\lfloor \cdot \rfloor$, and $x^{(j)}$ ceiling function, floor function, and $j$-th derivative of $x$ are represented.

II. PRELIMINARIES

In this section, some basic concepts of fractional-order calculus and useful definitions and lemmas are presented.

The following Caputo fractional derivative definition is adopted throughout this paper, due to its integer–order initial value that makes it applicable to physical systems.

$$_a^C D_t^\alpha = \frac{1}{\Gamma(m-\alpha)} \int_a^t (t-\tau)^{m-a-1} \left(\frac{d}{d\tau}\right)^m f(\tau)d\tau,$$

where $\Gamma(\cdot)$ is Gamma function defined by $\Gamma(\epsilon) = \int_0^\infty e^{-t}t^{\epsilon-1}dt$ and $m$ is the smallest integer that is equal to or greater than $\alpha$. In the sequel, wherever the notation $D^\alpha$ is used, the Caputo fractional derivative operator $_a^C D_t^\alpha$ is meant.

For the proofs of the theorems in this paper, we will use the following lemmas.

**Lemma 1** [18]. Let $f \in C^k[a,b]$ for some $a < b$ and some $k \in \mathbb{N}$. Moreover, let $n, \varepsilon > 0$ be such that there exists some $l \in N$ with $l \leq k$ and $n, n + \varepsilon \in [l-1, l]$. Then

$$_a^C D_t^\varepsilon {_a^C D_t^n} f = {_a^C D_t^{n+\varepsilon}} f. \tag{1}$$

**Lemma 2** [18]. Let $n > 0$, $n \notin \mathbb{N}$ and $m = \lceil n \rceil$. Moreover, assume that $f \in C^m[a,b]$. Then, $_a^C D_t^n f \in C^0[a,b]$ and $_a^C D_t^n f(a) = 0$.

**Lemma 3** [31]. Commensurate system $D^\alpha x(t) = Ax(t)$ is stable if the following conditions are satisfied

$$|\arg(eig(A))| > \alpha\pi/2, \tag{2}$$

for all eigenvalues $eig(A)$ of the matrix $A$.

**Lemma 4** [28]. Let $A \in \mathcal{R}^{n \times n}$, $0 < \alpha < 1$ and $\theta = (1-\alpha)\pi/2$. The fractional-order system $D^\alpha x(t) = Ax(t)$ is asymptotically stable if and only if there exists a positive definite Hermitian matrix $X = X^* > 0$, $X \in C^{n \times n}$ such that

$$(rX + \bar{r}\bar{X})^T A^T + A(rX + \bar{r}\bar{X}) < 0, \tag{3}$$

where $r = e^{\theta i}$.

**Definition 1.** For any square matrix $A = (a_{ij}) \in \mathcal{R}^{n \times n}$ and the set $\alpha_i$, $i = 1, \ldots, n$ the linear matrix function $F$ is defined as follows:

$$F(A, n, \alpha_i) = \left(\begin{cases} \begin{pmatrix} 0 & 1 & 0 & \cdots & 0 \\ 0 & 0 & 1 & \ddots & \vdots \\ \vdots & \vdots & \ddots & \ddots & 0 \\ 0 & 0 & \cdots & 0 & 1 \\ a_{ij} & 0 & 0 & \cdots & 0 \end{pmatrix}_{p_i \times p_j}, & i = j \\ \begin{pmatrix} 0 & 0 & \cdots & 0 \\ 0 & 0 & \cdots & 0 \\ \vdots & \vdots & \ddots & \vdots \\ a_{ij} & 0 & \cdots & 0 \end{pmatrix}_{p_i \times p_j}, & i \neq j \end{cases}\right) \quad i,j = 1, \ldots, n, \tag{4}$$

in which $p_i = \alpha_i/\alpha_C$ with $\alpha_C = \gcd(\alpha_i)$.

III. PROBLEM FORMULATION

Consider the following FO-LTI system

$$\begin{bmatrix} D^{\alpha_1}x_1(t) \\ \vdots \\ D^{\alpha_n}x_n(t) \end{bmatrix} = AX(t) + Bu(t), \quad 0 < \alpha_i < 2, \ i = 1, \ldots, n, \tag{5}$$

$$y(t) = CX(t),$$

subject to the initial conditions

$$X^{(j)}(0) = \left(x_{0i}^{(j)}\right)^T \in R^n, j = 0, \ldots, \lfloor \alpha_i \rfloor, \tag{6}$$

in which $X = (x_1 \ x_2 \ \cdots \ x_n)^T \in \mathcal{R}^n$ denotes the pseudo-state vector, $\alpha_i, i = 1, \ldots, n$ are the different fractional orders of the system, $u \in \mathcal{R}^l$ is the control input, and $y \in \mathcal{R}^m$ is the output vector. Furthermore, $A = (a_{ij}) \in \mathcal{R}^{n \times n}$, $B = (b_{ij}) \in \mathcal{R}^{n \times l}$, and $C = (c_{ij}) \in \mathcal{R}^{m \times n}$ are known constant matrices with appropriate dimensions.

Defining $\alpha_C = \gcd(\alpha_i)$, $p_i = \alpha_i/\alpha_C$, and $N = \sum_{i=1}^{n} p_i, i = 1, \ldots, n$, one can obtain the following equivalent pseudo-state space system

$$D^{\alpha_C} Z(t) = \mathbb{A} Z(t) + \mathbb{B} u(t), \tag{7}$$
$$y(t) = \mathbb{C} Z(t),$$

in which

$$Z = (z_{11} \ \cdots \ z_{1p_1} \ z_{21} \ \cdots \ z_{2p_2} \ \cdots \ z_{n1} \ \cdots \ z_{np_n})^T, \tag{8}$$

and

$$z_{ij} = D^{(j-1).\alpha_C} x_i, \ i = 1, \ldots, n, j = 1, \ldots, p_i, \tag{9}$$

and initial conditions are as follows

$$Z(0) = \left(z_{0ik} = \begin{cases} x_{0i} & k = 1 \\ x_{0i}^{(1)} & k\alpha_C = 1 \\ 0 & else \end{cases}\right) \in \mathcal{R}^N \quad \begin{matrix} i = 1, \ldots, n \\ k = 1, \ldots, p_i \\ j = 0, \ldots, \lfloor \alpha_i \rfloor \end{matrix}, \tag{10}$$

$$Z^{(1)}(0) = \left(z_{0ik}^{(1)} = \begin{cases} x_{0i}^{(1)} & k = 1 \\ 0 & else \end{cases}\right) \in \mathcal{R}^N$$

and $\mathbb{A}$ is easily obtained using the linear matrix function F in Definition 1, as follows

$$\mathbb{A} = F(A, n, \alpha_i) = (\mathbf{a}_{ij}) \in \mathcal{R}^{N \times N}, \quad i, j = 1, \ldots, n, \tag{11}$$

also we have

$$\mathbb{B} = (\mathbf{b}_{ij}) \in \mathcal{R}^{N \times l}, \quad \mathbb{C} = (\mathbf{c}_{ij}) \in \mathcal{R}^{m \times N}$$
$$\mathbf{b}_{ij} = \begin{bmatrix} 0 \\ B_{i,*} \end{bmatrix}_{p_i \times l}, \ (\mathbf{c}_{ij}) = [C_{*,j} \ \ 0]_{m \times p_i} \ i, j = 1, \ldots, n, \tag{12}$$

where $\mathbf{a}_{ij}$, $\mathbf{b}_{ij}$, and $\mathbf{c}_{ij}$ are submatrices in the $i$-th row and $j$-th column of partitioned matrices $\mathbb{A}$, $\mathbb{B}$, and $\mathbb{C}$.

IV. MAIN RESULTS

In this section, first, an equivalent single-order system for the given different-order system (5) is derived using the matrix function (4). Then, utilizing the equivalent system, a new stability condition is derived for system (5) and an LMI approach is proposed for designing a dynamic output feedback control law to stabilize the fractional-order system (5), using the stabilizing controller parameters of system (7).

*A. Equivalency*

First, equivalency of systems (5) and (7) is proved in the following theorem.

**Theorem 1**. Consider the fractional-order linear system (5). Defining $\alpha_C$ to be the greatest common divisor of $\alpha_i$s, the FO-LTI systems (5) and (7) subject to the initial conditions $X(0)$ and $Z(0)$ are equivalent in the following sense:
  1. Whenever $Z$, defined in (7) and (8), with $z_{i1} \in C^{\lceil \alpha_i \rceil}[0, b]$ for some $b > 0$ is the pseudo-state of the FO-LTI system (7), $X$ is the pseudo-state of (5).

2. Whenever $X$ is the pseudo-state of (5), $Z$, defined in (7) and (8), is the pseudo-state of the FO-LTI system (7).

**Proof.** In the case that all fractional derivatives of FO-LTI system (5) are equal, we have $\alpha_1 = \cdots = \alpha_n = \alpha_C$ which brings about the equality of systems (5) and (7). Therefore, from now on we assume that at least one of the $\alpha_i$s is different from the others. As a result we have $0 < \alpha_C < 1$.

In order to prove the first claim, we have to assume that $Z = (z_{11} \quad \cdots \quad z_{1p_1} \quad z_{21} \quad \cdots \quad z_{2p_2} \quad \cdots \quad z_{n1} \quad \cdots \quad z_{np_n})^T$ is the pseudo-state vector of system (7), and we define $x_i = z_{i1}, i = 1, \dots, n$. Since $\alpha_C < 1$, by a repeated application of Lemma 1 combined with (9), we obtain

$$D^{\alpha_C} x_i = D^{\alpha_C} z_{1i} = z_{2i},$$
$$D^{2\alpha_C} x_i = D^{\alpha_C} D^{\alpha_C} z_{1i} = D^{\alpha_C} z_{2i} = z_{3i}, \tag{13}$$
$$\vdots$$
$$D^{p_i \alpha_C} x_i = D^{\alpha_C} D^{(p_i-1)\alpha_C} z_{1i} = D^{\alpha_C} z_{(p_i-1)i} = \mathbb{A}_{i,*} Z(t) + \mathbb{B}_{i,*} u(t), \quad i = 1, \dots, n.$$

According to the rows of $\mathbb{A}$ and $\mathbb{B}$, which are made from $A$ and $B$ rows respectively, and the fact that $z_{i1} = x_i$, the last equation can be rewritten as follows

$$D^{p_i \alpha_C} x_i = A_{i,*} X(t) + B_{i,*} u(t). \tag{14}$$

By definition, $p_i \alpha_C = (\alpha_i/\alpha_C)\alpha_C = \alpha_i$ and considering (14) for $i = 1, \dots, n$ we have

$$\begin{bmatrix} D^{\alpha_1} x_1(t) \\ \vdots \\ D^{\alpha_n} x_n(t) \end{bmatrix} = AX(t) + Bu(t).$$

Therefore, $X(t)$ is the pseudo-state vector of system (5). In addition, it is obvious from equation system (13) that for $j = 1, \dots, p_i$ we have $x_i^{(j)}(0) = x_{0i}^{(j)} = D^j x_i(0) = z_{i1}^{(j)}(0) = z_{0i1}^{(j)}$. Therefore, $X(t)$ satisfies the initial conditions (6).

For the second claim we have to assume that $X = (x_1 \quad x_2 \quad \cdots \quad x_n)^T$ is the pseudo-state vector of system (5) with the initial conditions (6). The equation system (13) is valid in this case too, and therefore it follows that $Z$ is the pseudo-state vector of system (7) with the initial conditions $Z^{(j)}(0) = \left(z_{0ik}^{(j)}\right) = \left(x_{0i}^{(j)}\right), j = j = 0, \dots, \lfloor \alpha_i \rfloor$ whenever $k = 1$ and $Z(0) = (z_{0ik}) = x_{0i}^{(1)}$ for $k\alpha_C = 1$. Finally, an application of Lemma 2 reveals that $Z^{(j)}(0) = \left(z_{0ik}^{(j)}\right) = 0$ in the other cases, and it completes the proof. ∎

*B. Stability*

In this subsection a necessary and sufficient condition is established for the asymptotic stability of system (5) with $u(t) \equiv 0$.

**Theorem 2.** The fractional-order linear system (5) with $u(t) \equiv 0$ is asymptotically stable for different orders $\alpha_i$, if and only if there exists a Hermitian matrix $X \in \mathcal{C}^{N \times N}$ such that

$$(rX + \bar{r}\bar{X})^T \mathbb{A}^T + \mathbb{A}(rX + \bar{r}\bar{X}) < 0, \tag{15}$$

in which $r = e^{\theta i}$ with $\theta = (1 - \alpha_C)\pi/2$ holds, and $\mathbb{A}$ is defined in (11).

**Proof.** It can be easily concluded that since the fractional orders $0 < \alpha_i < 2$, $i = 1, \dots, n$ are different from each other, then $0 < \alpha_C < 1$. Therefore, It follows from Lemma 4 that the fractional-order system (7) is asymptotically stable if and only if there exists a positive definite Hermitian matrix $X = X^*, X \in \mathcal{C}^{N \times N}$ such that (15) holds, and since, according to Theorem 1, the pseudo-states of original system (5) are in the subset of system (7) pseudo-states, asymptotic stability of (7) results that of the system (5).

For the proof of necessity we suppose that system (5) is asymptotically stable and system (7) is not asymptotically stable, that means there is at least a $z_{ij}(t), j \neq 1$ that does not tend to zero as $t \to \infty$. So it is a divergent or an oscillatory signal. According to the structure of matrix $\mathbb{A}$, since $D^{\alpha_C} z_{i(j-1)} =$

$z_{ij}(t), j \neq 1$, $z_{i(j-1)}(t), j \neq 1$ is correspondingly a divergent or an oscillatory signal. Repeating this procedure concludes the instability of $z_{i1}(t) = x_i(t)$, which contradicts with the initial assumption of stability of system (5), and it completes the proof. ∎

## C. Stabilization

The main purpose of this subsection is to design a fixed-order dynamic output feedback controller that asymptotically stabilizes FO-LTI system (5). Hence, the following controller is presented

$$D^{\alpha_C} X_C(t) = A_C X_C(t) + B_C y(t), \quad (16)$$
$$u(t) = C_C X_C(t) + D_C y(t),$$

with $X_C(t) = \begin{pmatrix} x_{C_1}(t) & x_{C_1}(t) & \cdots & x_{C_{n_C}}(t) \end{pmatrix}^T \in \mathcal{R}^{n_C}$, in which $n_C$ is the arbitrary order of the controller, and $A_C, B_C, C_C$, and $D_C$ are the controller unknown parameters with appropriate dimensions. The resulting closed-loop augmented FO-LTI system using (5) and (16) is as follows

$$\begin{bmatrix} D^{\alpha_1} x_1(t) \\ \vdots \\ D^{\alpha_n} x_n(t) \\ D^{\alpha_C} x_{C_1}(t) \\ \vdots \\ D^{\alpha_C} x_{C_{n_C}}(t) \end{bmatrix} = A_{Cl} X_{Cl}(t), \quad \begin{array}{l} 0 < \alpha_i < 2, \\ \alpha_C = \gcd(\alpha_i) \end{array} \quad i = 1, \ldots, n, \quad (17)$$

where

$$X_{Cl}(t) = \begin{bmatrix} X(t) \\ X_C(t) \end{bmatrix}, \quad A_{Cl} = \begin{bmatrix} A + B D_C C & B C_C \\ B_C C & A_C \end{bmatrix}. \quad (18)$$

Next, a stabilization result is established.

**Theorem 3**. Considering closed-loop system in (17), if a positive definite Hermitian matrix $P = P^* \in \mathcal{C}^{(N+n_C) \times (N+n_C)}$ in the form of

$$P = diag(P_S, P_C), \quad P_S \in \mathcal{C}^{N \times N}, P_C \in \mathcal{C}^{n_C \times n_C}, \quad (19)$$

alongside with matrices $W_i$, $i = 1, \ldots, 4$ exist in a way that

$$\begin{bmatrix} \mathbb{A}(rP_S + \bar{r}\overline{P_S}) + (rP_S + \bar{r}\overline{P_S})^T \mathbb{A}^T + \mathbb{B} W_4 + W_4^T \mathbb{B}^T, & \mathbb{B} W_3 + W_2^T \\ W_2 + W_3^T \mathbb{B}^T & W_1 + W_1^T \end{bmatrix} < 0, \quad (20)$$

where $r = e^{\theta i}$ with $\theta = (1 - \alpha_C)\pi/2$ holds, then, the dynamic output feedback controller parameters of

$$A_C = W_1(rP_C + \bar{r}\overline{P_C})^{-1}, \quad B_C = W_2(rP_S + \bar{r}\overline{P_S})^{-1} \mathbb{C}^\uparrow, \quad (21)$$
$$C_C = W_3(rP_C + \bar{r}\overline{P_C})^{-1}, \quad D_C = W_4(rP_S + \bar{r}\overline{P_S})^{-1} \mathbb{C}^\uparrow,$$

make the FO-LTI system in (5) asymptotically stable.

**Proof**. It follows from Theorem 1 that the closed-loop FO-LTI system (17) is asymptotically stable if there exists a positive definite Hermitian matrix $P = P^*, P \in \mathcal{C}^{(N+n_C) \times (N+n_C)}$ such that

$$(rP + \bar{r}\bar{P})^T \mathbb{A}_{Cl}^T + \mathbb{A}_{Cl}(rP + \bar{r}\bar{P}) < 0, \quad (22)$$

where

$$\mathbb{A}_{Cl} = F(A_{Cl}, N + n_C, \{\alpha_1, \ldots, \alpha_n, \alpha_C, \ldots, \alpha_C\}), \quad (23)$$

in which $F$ is defined in (4). By some manipulations, it can be easily concluded that

$$F(A_{Cl}, N + n_C, \{\alpha_1, \ldots, \alpha_n, \alpha_C, \ldots, \alpha_C\}) = \begin{bmatrix} \mathbb{A} + \mathbb{B}D_C\mathbb{C} & \mathbb{B}C_C \\ B_C\mathbb{C} & A_C \end{bmatrix}, \quad (24)$$

where $\mathbb{A}$, $\mathbb{B}$, and $\mathbb{C}$ are defined in (11) and (12). According to (22) and (24), and assuming matrix $P = P^* \in \mathcal{C}^{(N+n_C) \times (N+n_C)}$ in the form of (19), the inequality (22) can be written as follows

$$\begin{bmatrix} E_{11} & E_{12} \\ E_{21} & E_{22} \end{bmatrix} < 0, E_{11} = \mathbb{A}(rP_S + \bar{r}\overline{P_S}) + (rP_S + \bar{r}\overline{P_S})^T \mathbb{A}^T + \mathbb{B}D_C\mathbb{C}(rP_S + \bar{r}\overline{P_S}) + (rP_S + \bar{r}\overline{P_S})^T \mathbb{C}^T D_C^T \mathbb{B}^T, E_{12} = \mathbb{B}C_C(rP_C + \bar{r}\overline{P_C}) + (rP_S + \bar{r}\overline{P_S})^T \mathbb{C}^T B_C^T, \quad (25)$$

$$E_{21} = B_C\mathbb{C}(rP_S + \bar{r}\overline{P_S}) + (rP_C + \bar{r}\overline{P_C})^T C_C^T \mathbb{B}^T, E_{22} = A_C(rP_C + \bar{r}\overline{P_C}) + (rP_C + \bar{r}\overline{P_C})^T A_C^T.$$

The matrix inequality (25) is not linear due to several multiplications of variables. Therefore, by linearizing change of variables as

$$W_1 = A_C(rP_C + \bar{r}\overline{P_C}), W_2 = B_C C(rP_S + \bar{r}\overline{P_S}), W_3 = C_C(rP_C + \bar{r}\overline{P_C}), W_4 = D_C C(rP_S + \bar{r}\overline{P_S}), \quad (26)$$

$E_{ij}, i, j = 1,2$ in (25) can be rewritten as

$$E_{11} = \mathbb{A}(rP_S + \bar{r}\overline{P_S}) + (rP_S + \bar{r}\overline{P_S})^T \mathbb{A}^T + \mathbb{B}W_4 + W_4^T \mathbb{B}^T, \quad (27)$$
$$E_{12} = \mathbb{B}W_3 + W_2^T, E_{21} = W_2 + W_3^T \mathbb{B}^T, E_{22} = W_1 + W_1^T.$$

which makes (25) equivalent to inequality (20), and it completes the proof. ∎

**Remark 1.** The stability checking method proposed in Theorem 2 is based on the greatest common divisor of plant orders however; available method in [17] is based on the least common multiple of the denominators of plant orders, which often leads to high-order characteristic equations. Therefore, only for some specific cases checking the stability of different-order FO-LTI system (5), is of same degree of difficulty for both methods and otherwise our proposed method is easier to check.

**Remark 2.** In Theorem 3, control parameters are obtained by finding a suitable block-diagonal positive definite matrix $P$, which can potentially result in conservatism of the problem.

V. SIMULATION

In this section, some numerical examples are given to validate the effectiveness of the derived results. In this paper, we use YALMIP parser [32] and SeDuMi [33] solver in Matlab tool [34].

*A. Example 1*

Consider system (5) with the following parameters

$$A = \begin{bmatrix} 1 & 0 & 1.5 \\ 1 & -2 & 0.5 \\ 1 & 1 & -3 \end{bmatrix}, \alpha_1 = 0.93, \alpha_2 = 1.55, \alpha_3 = 1.24. \quad (28)$$

According to the stability criterion proposed in [19], characteristic equation of above system is as follows

$$\lambda^{372} - \lambda^{279} + 3\lambda^{248} + 2\lambda^{217} - 4.5\lambda^{155} - 2\lambda^{124} + 5.5\lambda^{93} - 10 = 0. \quad (29)$$

The roots of characteristic equation (29) along with the stability boundary $\pm \gamma \pi/2$ are depicted in Fig. 1. It can be concluded from [17] that the system (28) is unstable since not all of the roots of equation (29) satisfy the condition $|\arg(\lambda)| > \gamma \pi/2$, in which $\gamma = 0.01$. The same conclusion can be drawn from proposed Theorem 2 with a relatively simpler method. The eigenvalues of $\mathbb{A}$ and stability boundaries $\pm \alpha_C \pi/2$, with $\alpha_C = 0.31$ are demonstrated in Fig. 2. According to Lemma 3 and Fig. 2, the system $D^{\alpha_C} x(t) = \mathbb{A}x(t)$ is unstable for $\mathbb{A} = F(A, 3, (\alpha_1, \alpha_2, \alpha_3))$, since not all of the eigenvalues of $\mathbb{A}$ are located on the left side of $\pm\alpha_c\pi/2$ boundaries. Subsequently, according to Theorem 2, system (5) with the parameters in (28) is unstable.

Since the characteristic equation (29) is of order 372 and $N = p_1 + p_2 + p_3 = 12$, checking the arguments of eigenvalues of $\mathbb{A}$ is far easier than checking the arguments of roots of characteristic equation (29).

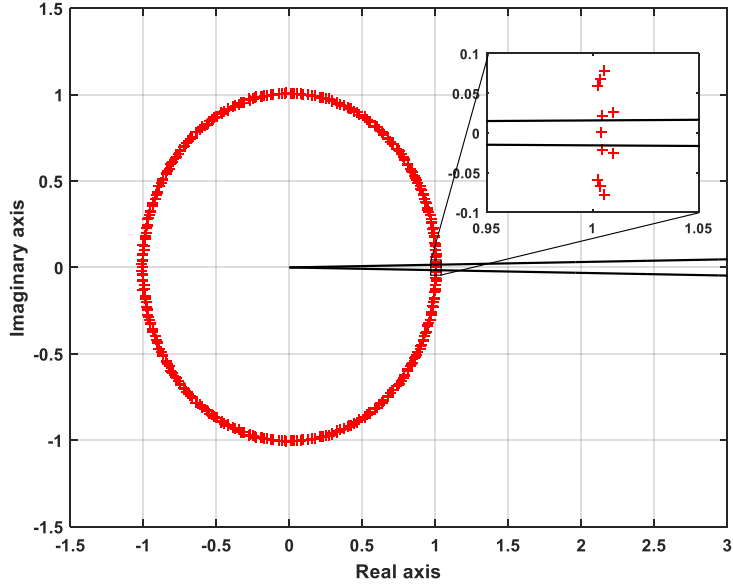

Fig. 1. Roots of characteristic equation (29) and stability boundaries $\pm\gamma\pi/2$ with $\gamma = 0.01$ for the system of Example 1.

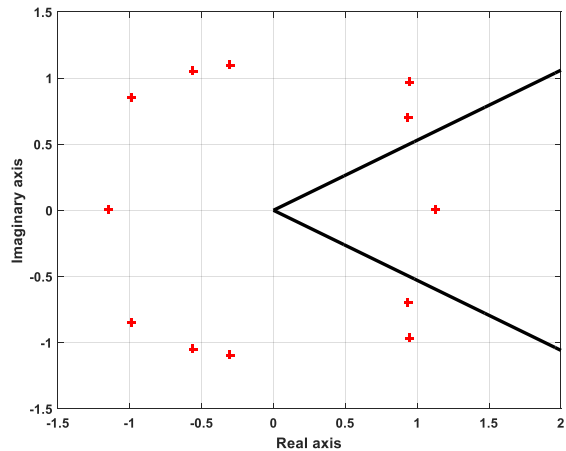

Fig. 2. The eigenvalues of $\mathbb{A}$ and stability boundaries $\pm\alpha_c\pi/2$, with $\alpha_c = 0.31$ for the system of Example 1.

## B. Example 2

Sallen–Key filters are one of the most common and well-known filters especially because of their simplicity [35]. The conventional Sallen–Key family is used to implement second-order active filters by utilizing two integer-order capacitors and one operational amplifier (op-amp) [36]. In addition, because of the importance of these filter family, fractional-order type of them is also generalized in the presence of the extra fractional-order parameters The transfer function of the fractional-order Sallen–Key filter illustrated in Fig. 3 is given as follows [36]

$$\frac{V_{out}}{V_{in}} = \frac{\frac{1}{C_1 C_2 R_1 R_2}}{s^{\alpha_1+\alpha_2} + \frac{1-R_4/R_3}{R_2 C_2}s^{\alpha_1} + \left(\frac{1}{R_2 C_1} + \frac{1}{R_1 C_1}\right)s^{\alpha_2} + \frac{1}{C_1 C_2 R_1 R_2}}, \tag{30}$$

where $0 < \alpha_1, \alpha_2 < 2$ are the fractional orders of capacitors.

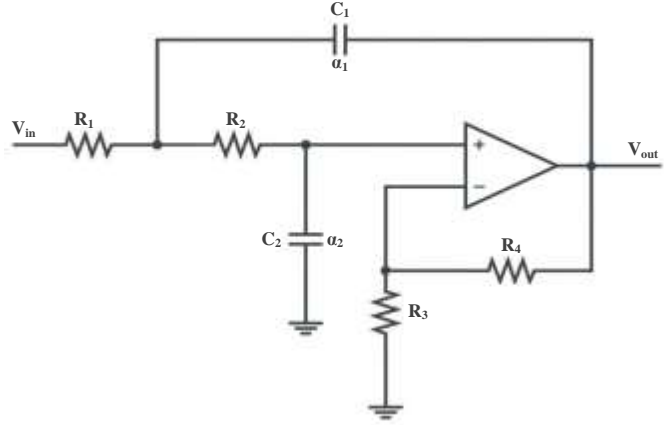

Fig. 3. Fractional order Sallen–Key filter [36].

The stability analysis of Sallen–Key fractional-order filters was investigated in [36] by drawing stability contour for the filter at different combinations of the characteristic equation parameters. In this example the Sallen–Key filter stability with two fractional-order capacitors of different orders $(\alpha_1, \alpha_2)$ is checked using proposed Theorem 2. Numerical values of elements of fractional-order Sallen–Key filter are considered as $R_1 = 7.11 k\Omega$, $R_2 = 30.63 k\Omega$, $R_4/R_3 = 0.8836$, $C_1 = 20\mu F$, $C_2 = 50\mu F$, and $(\alpha_1, \alpha_2) = (0.78, 1.17)$, using which the pseudo-state space system (5) with the following parameters can be obtained.

$$A = \begin{bmatrix} -8.6647 & -7.0323 \\ 4.1489 & -0.0760 \end{bmatrix}, B = \begin{bmatrix} 1 \\ 0 \end{bmatrix}, C = \begin{bmatrix} 0 & 4591.8 \end{bmatrix}, \alpha_1 = 0.78, \alpha_2 = 1.17. \tag{31}$$

According to the stability criterion in [17], characteristic equation of above system is as follows

$$\lambda^{195} + 8.7\lambda^{117} + 0.0760\lambda^{78} + 28.5182 = 0. \tag{32}$$

The roots of characteristic equation (32) and the stability boundary $\pm\gamma\pi/2$ are depicted in Fig. 4. It can be concluded from [17] that the system (31) is asymptotically stable since roots of equation (32) satisfy the condition $|\arg(\lambda)| > \gamma\pi/2$, in which $\gamma = 0.01$. The same conclusion can be drawn from proposed Theorem 2 The eigenvalues of $\mathbb{A}$ and stability boundaries $\pm\alpha_C\pi/2$, with $\alpha_C = 0.39$ are demonstrated in Fig. 5. According to Lemma 3 and Fig. 5, the system $D^{\alpha_C}x(t) = \mathbb{A}x(t)$ is stable for $\mathbb{A} = F(A, 2, (\alpha_1, \alpha_2))$. Subsequently, according to Theorem 2, system (5) with the parameters in (31) is asymptotically stable. Since the characteristic equation (32) is of order 195 and $N = p_1 + p_2 = 5$, checking the arguments of eigenvalues of $\mathbb{A}$ is far easier than checking the arguments of roots of characteristic equation (32), which is obvious from Figures 4 and 5.

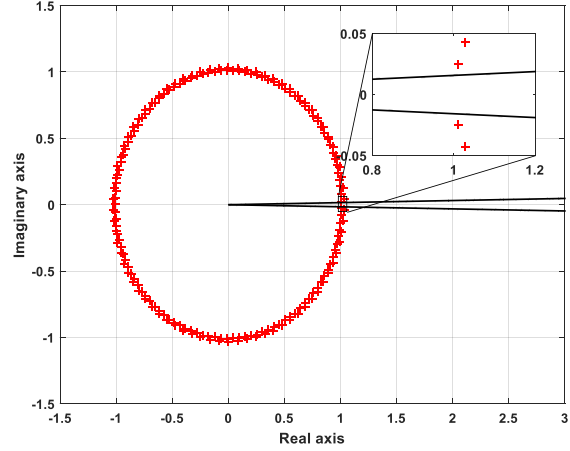

Fig. 4. Roots of characteristic equation (32) and stability boundaries $\pm\gamma\pi/2$ with $\gamma = 0.01$ for the system of Example 2.

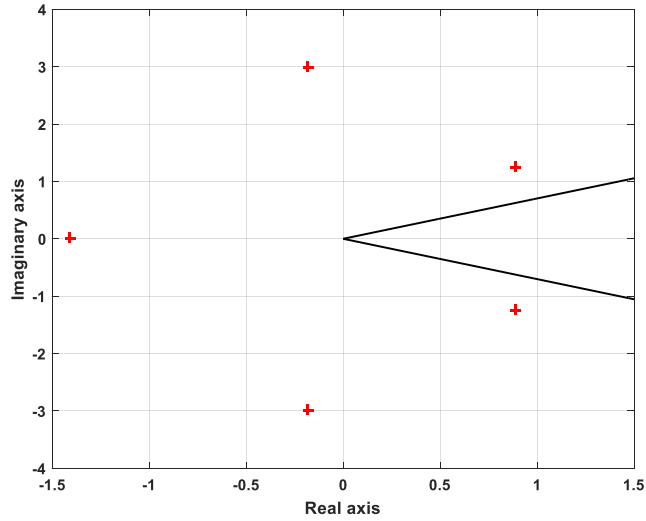

Fig. 5. The eigenvalues of $\mathbb{A}$ and stability boundaries $\pm\alpha_c\pi/2$, with $\alpha_c = 0.39$ for the system of Example 2.

## C. Example 3

Dynamic output feedback stabilization problem of the fractional-order system (5) is considered with the following parameters

$$A = \begin{bmatrix} 3 & 1 \\ -1 & -2 \end{bmatrix}, \ B = \begin{bmatrix} 3 \\ 2 \end{bmatrix}, \ C = [-2 \ \ 0], \alpha_1 = 0.6, \alpha_2 = 1.5. \tag{33}$$

The eigenvalues of $\mathbb{A}$ and stability boundaries $\pm\alpha_c\pi/2$, with $\alpha_c = 0.3$ are demonstrated in Fig. 6. According to Lemma 3 and Fig. 6, the system $D^{\alpha_c}x(t) = \mathbb{A}x(t)$ is unstable for $\mathbb{A} = F(A, 2, (0.6, 1.5))$, since not all of the eigenvalues of $\mathbb{A}$ are located on the left side of $\pm\alpha_c\pi/2$ boundaries. Thus, according to Theorem 2, system (5) with the parameters in (33) is unstable. However, according to Theorem 3, it can be concluded that this fractional-order system is asymptotically stabilizable utilizing the obtained dynamic output feedback controllers of arbitrary orders in the form of (16), tabulated in Table 1. The eigenvalues of $\mathbb{A}_{Cl}$ are located in stability region which is depicted in Fig. 6. According to Theorem 3, stability of $\mathbb{A}_{Cl}$ implies the stability of $A_{Cl}$. Fig. 7 shows the time response of open-loop system and the resulted closed-loop system of form (17) via obtained controllers with $n_c = 0$ and 1, from which it is obvious that all of the closed-loop system pseudo-states asymptotically converge to zero. It can be concluded that, the obtained dynamic output feedback controllers, have more appropriate control actions compared to that of the static controller of $n_c = 0$.

TABLE 1. CONTROLLER PARAMETERS OF EXAMPLE 3 OBTAINED BY THEOREM 3

| $n_c$ | $A_c$ | $B_c$ | $C_c$ | $D_c$ |
|---|---|---|---|---|
| 0 | 0 | 0 | 0 | 1.28 |
| 1 | -21.78 | 0.053 | 2.46 | 1.20 |
| 2 | $\begin{bmatrix} -14.41 & -3.31 \\ -3.31 & -15.35 \end{bmatrix}$ | $\begin{bmatrix} 0.081 \\ 0.094 \end{bmatrix}$ | $[\;4.01\;\;\;4.63]$ | 1.14 |

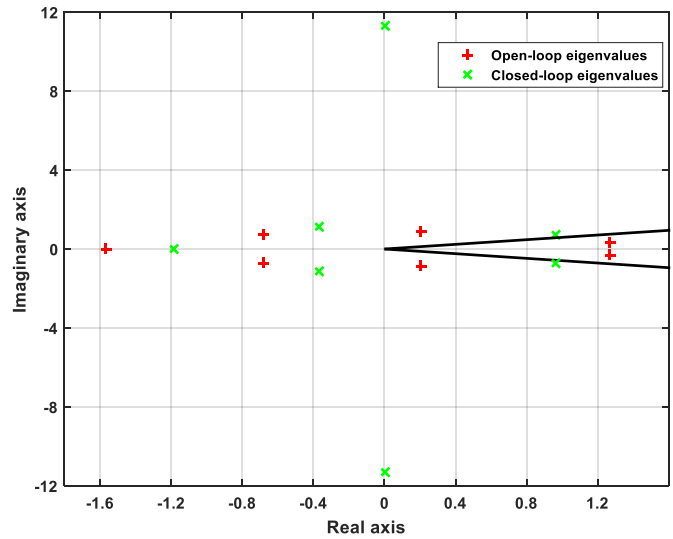

Fig. 6. The eigenvalues of $\mathbb{A}$, $\mathbb{A}_{Cl}$, and stability boundaries $\pm \alpha_c \pi/2$, with $\alpha_c = 0.3$ for the system of Example 3 with $n_c = 0$.

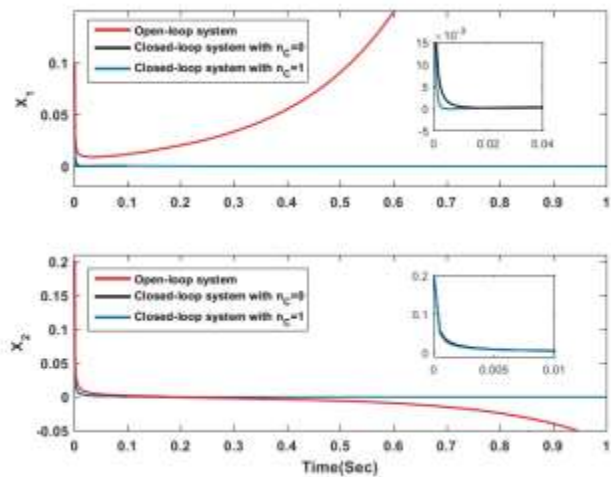

Fig. 7. The time response of the open-loop (red) and closed-loop system in Example 3 via obtained output feedback controllers with $n_c = 0$ (black), $n_c = 1$ (blue).

## VI. Conclusion

In this paper using an appropriate linear matrix function, a single-order equivalent system for the given commensurate- different-order system is introduced through which a new stability condition has been obtained that is easier to check in practice compared with the conditions known up to now (Theorem 1 and 2). Then the stabilization problem of FO-LTI systems with different fractional orders via a dynamic output feedback controller with a predetermined order has been addressed based on the proposed stability criterion (Theorem 3). Eventually, some numerical examples are presented to show the validity of the proposed theorems.

## References

1. Lu, J.-G., and Y.Chen, "Stability and stabilization of fractional-order linear systems with convex polytopic uncertainties," *Fract. Calc. Appl. Anal.* **16,** 142–157 (2012).
2. Badri, V., and M. S. Tavazoei, "Fractional order control of thermal systems: achievability of frequency-domain requirements," *Nonlinear Dyn.* **80,** 1773–1783 (2014).
3. Badri, V., and M. S.Tavazoei, "On tuning fractional order [proportional–derivative] controllers for a class of fractional order systems," *Automatica* **49,** 2297–2301 (2013).
4. Badri, V. and M. S.Tavazoei, "Some Analytical Results on Tuning Fractional-Order [Proportional-Integral] Controllers for Fractional-Order Systems," *IEEE Trans. Control Syst. Technol.* **24,** 1059–1066 (2016).
5. Rajagopal, K., and F. Nazarimehr, and A. Karthikeyan, and A. Srinivasan, and S. Jafari, "Fractional Order Synchronous Reluctance Motor: Analysis, Chaos Control and FPGA Implementation," *Asian J. Control* (2018) doi:10.1002/asjc.1690
6. Liu, X., Zhang, Z. & Liu, H. Consensus Control of Fractional-Order Systems Based on Delayed State Fractional Order Derivative. Asian J. Control 19, 2199–2210 (2017).
7. Podlubny, I. "Fractional-order systems and PI/sup /spl lambda//D/sup /spl mu//-controllers," *IEEE Trans. Autom. Control* **44,** 208–214 (1999).
8. Jin, Y., Chen, Y. Q. and D. Xue, "Time-constant robust analysis of a fractional order [proportional derivative] controller," *IET Control Theory Appl.* **5,** 164–172 (2011).
9. Badri, P., Sojoodi, M. "Robust Fixed-order Dynamic Output Feedback Controller Design for Fractional-order Systems," *IET Control Theory Amp Appl.* (2018). doi:10.1049/iet-cta.2017.0608
10. Kaczorek, T. "Positive Linear Systems Consisting of n Subsystems With Different Fractional Orders," *IEEE Trans. Circuits Syst. Regul. Pap.* **58,** 1203–1210 (2011).
11. Wei, Y., and Y. Hu, and L. Song, and Y. Wang, "Tracking Differentiator Based Fractional Order Model Reference Adaptive Control: The $1 < \alpha < 2$ Case," in *53rd IEEE Conference on Decision and Control* 6902–6907 (2014).
12. Taghavian, H. and M. S. Tavazoei, "Robust stability analysis of uncertain multiorder fractional systems: Young and Jensen inequalities approach," *Int. J. Robust Nonlinear Control* (2018) doi:10.1002/rnc.3919
13. Badri, V. and M. S. Tavazoei, "Achievable Performance Region for a Fractional-Order Proportional and Derivative Motion Controller," *IEEE Trans. Ind. Electron.* **62,** 7171–7180 (2015).
14. Badri, V. and Saleh M. S. Tavazoei, "Simultaneous Compensation of the Gain, Phase, and Phase-Slope" *J. Dyn. Syst. Meas. Control* **138,** 121002–121002–7 (2016).
15. Skaar, S. B., and A. N. Michel, and R. K. Miller, "Stability of viscoelastic control systems" *IEEE Trans. Autom. Control* **33,** 348–357 (1988).
16. Ammour, A. S., Djennoune, S., Aggoune, W. & Bettayeb, M. "Stabilization of Fractional-Order Linear Systems with State and Input Delay," *Asian J. Control* **17,** 1946–1954 (2015).
17. Deng, W., Li, C. & Lü, J. "Stability analysis of linear fractional differential system with multiple time delays," *Nonlinear Dyn.* **48,** 409–416 (2007).
18. Diethelm, K. *"The Analysis of Fractional Differential Equations: An Application-Oriented Exposition Using Differential Operators of Caputo Type*. (Springer, 2010).
19. Chen, L. *et al.* "Stabilization of Uncertain Multi-Order Fractional Systems Based on the Extended State Observer," *Asian J. Control* (2018) doi:10.1002/asjc.1618
20. Badri, P., Amini, A. & Sojoodi, M. "Robust fixed-order dynamic output feedback controller design for nonlinear uncertain suspension system," *Mech. Syst. Signal Process.* **80,** 137–151 (2016).
21. Amini, A., Azarbahram, A. & Sojoodi, M. $H_\infty$ Consensus of nonlinear multi-agent systems using dynamic output feedback controller: an LMI approach. Nonlinear Dyn. 85, 1865–1886 (2016).
22. Zhang, H., Wang, X. & Lin, X. Stability and Control of Fractional Chaotic Complex Networks with Mixed Interval Uncertainties. Asian J. Control 19, 106–115 (2017).

23. Dadras, S., Dadras, S. & Momeni, H. Linear Matrix Inequality Based Fractional Integral Sliding-Mode Control of Uncertain Fractional-Order Nonlinear Systems. J. Dyn. Syst. Meas. Control 139, 111003–111003–7 (2017).
24. Amini, A., Mohammadi, A. & Asif, A. "Event-based consensus for a class of heterogeneous multi-agent systems: An LMI approach," in *2017 IEEE International Conference on Acoustics, Speech and Signal Processing (ICASSP)* 3306–3310 (2017).
25. Thuan, M. V. & Huong, D. C. "New Results on Stabilization of Fractional-Order Nonlinear Systems via an LMI Approach," *Asian J. Control* (2018) doi:10.1002/asjc.1644
26. Ji Yude, Du Mingxing & Guo Yanping. Stabilization of Non-Linear Fractional-Order Uncertain Systems. Asian J. Control 20, 669–677 (2017).
27. Moze, M., Sabatier, J. & Oustaloup, A. "LMI Tools for Stability Analysis of Fractional Systems," 1611–1619 (2005). doi:10.1115/DETC2005-85182
28. Farges, C., Moze, M. & Sabatier, J. "Pseudo-state feedback stabilization of commensurate fractional order systems," *Automatica* **46,** 1730–1734 (2010).
29. Lu, J.-G. & Chen, G. "Robust Stability and Stabilization of Fractional-Order Interval Systems: An LMI Approach," *IEEE Trans. Autom. Control* **54,** 1294–1299 (2009).
30. Sontag, E. D. "*Mathematical Control Theory: Deterministic Finite Dimensional Systems*," (Springer Science & Business Media, 2013).
31. Matignon, D. "Stability properties for generalized fractional differential systems," *ESAIM Proc.* **5,** 145–158 (1998).
32. Löfberg, J. "YALMIP : a toolbox for modeling and optimization in MATLAB," in *2004 IEEE International Symposium on Computer Aided Control Systems Design* 284–289 (2004). doi:10.1109/CACSD.2004.1393890
33. Labit, Y., Peaucelle, D. & Henrion, D. "SEDUMI INTERFACE 1.02: a tool for solving LMI problems with SEDUMI. in *Proceedings*," IEEE International Symposium on Computer Aided Control System Design 272–277 (2002).
34. Higham, D. J. & Higham, N. J. *MATLAB Guide: Second Edition*. (SIAM, 2005).
35. Sallen, R. P. & Key, E. L. "A practical method of designing RC active filters," *IRE Trans. Circuit Theory* **2,** 74–85 (1955).
36. Soltan, A., Radwan, A. G. & Soliman, A. M. "Fractional Order Sallen–Key and KHN Filters: Stability and Poles Allocation," *Circuits Syst. Signal Process.* **34,** 1461–1480 (2015).